\documentstyle{article}

\def\SPH{SPH}
\def\fig{Fig.~}

\def\AaA{A\&A}

\def\AJ{AJ}

\def\ApJ{ApJ}

\def\ApJS{ApJS}

\def\ARAA{ARA\&A}

\def\MN{MNRAS}

\def\PhD{PhD thesis}
\def\Pre{Preprint}

\def\etal{{\it et al.\thinspace}}
\def\eg{{\it e.g.\ }}

\def\spose#1{\hbox to 0pt{#1\hss}}
\def\approxlt{\mathrel{\spose{\lower 3pt\hbox{$\sim$}}
	\raise 2.0pt\hbox{$<$}}}
\def\approxgt{\mathrel{\spose{\lower 3pt\hbox{$\sim$}}
	\raise 2.0pt\hbox{$>$}}}

\def\<{\thinspace}

\def\boxit#1{\vbox{\hrule\hbox{\vrule\kern3pt\vbox{\kern3pt
          #1 \kern3pt}\kern3pt\vrule}\hrule}}
\begin{document}
\vskip 1truein
\Large
\centerline{\bf An Owner's Guide to Smoothed Particle Hydrodynamics}
\normalsize
\vskip 1truein
\centerline{T.\ J.\ Martin, F.\ R.\ Pearce, P.\ A.\ Thomas}
\vskip 1truecm
\noindent {\it Astronomy Centre, Sussex University, Falmer, Brighton,
BN1 9QH.}
\vskip 1truecm
{\narrower
\noindent{\bf ABSTRACT}
We present a practical guide to Smoothed Particle Hydrodynamics (\SPH)
and its application to astrophysical problems.  Although remarkably
robust, \SPH\ must be used with care if the results are to be
meaningful since the accuracy of \SPH\ is sensitive to the arrangement
of the particles and the form of the smoothing kernel.  In particular,
the initial conditions for any \SPH\ simulation must consist of
particles in dynamic equilibrium.

We describe some of the numerical difficulties that may be encountered
when using \SPH, and how these may be overcome.  Through our
experience in using \SPH\ code to model convective stars, galaxy
clusters and large scale structure problems we have developed many
diagnostic tests.  We give these here as an aid to rapid
identification of errors, together with a list of basic prerequisites
for the most efficient implementation of
\SPH.

\vskip 1truecm
\noindent{\bf Key Words:} Hydrodynamics, Methods: numerical}
\section{Introduction}

Smoothed Particle Hydrodynamics, (\SPH), is an N-body integration
scheme introduced by Lucy (1977) and Gingold \& Monaghan
(1977) as an
attempt to model continuum physics avoiding the limitations of grid
based finite difference methods.  It has since been used in a wide
variety of astrophysical applications and is now a standard numerical
tool in many astrophysical problems.  Recent examples include the
formation of planetoids Benz \etal (1989) through investigations of
accretion disc instabilities (Lanzafume \etal 1993, Meglicki \etal
1993) and star
formation in collapsing molecular clouds (Pongracic \etal 1993),
to collisions between galaxies (Hernquist \& Katz 1989,
Steinmetz \& M\"{u}ller 1993),
clusters of galaxies (Pearce \etal 1993,
Navarro \& White 1993)
and the formation of galactic clusters (Thomas \& Couchman 1992,
Evrard 1990).

Most of the advantages of \SPH\ arise directly from its Lagrangian
nature.  There are no constraints imposed either on the geometry of
the system or in how far it may evolve from the initial conditions,
and the number density of the particles maps onto the mass density of
the fluid leading directly to increased resolution in high density
regions.  Furthermore, \SPH\ is a `low-level' computational scheme, in
that the physical equations are applied in their most basic, intuitive
form --- a consequence of always being able to specify the interaction
between individual particles.

However, although the physical interpretation may be simplified, the
numerical interpretation has become more complicated.  In place of a
fixed grid upon which differential equations are evaluated,
information about the fluid is now carried by an unconstrained set of
particles distributed by dynamical forces.  The interaction between
individual particles may be simple, but describing the the behaviour
of the collection is not.  It turns out that the distribution of the
particles is very important in an \SPH\ calculation.

A comprehensive theoretical introduction to and review of \SPH\ has
recently been given by Monaghan (1992).  In particular, he derives the
equations of motion in forms suitable for \SPH\ calculations and
explains which forms are most appropriate for different physical
applications.

In this paper we aim to give a practical introduction to \SPH.
We describe some of the problems that we have encountered while
using \SPH, and how these may be overcome.

In Section~2 we begin by describing where the idealised mathematical
formalism breaks down and comment on the importance of the particle
distribution.  We discuss the sources of systematic and statistical
error and how these depend upon the form of the smoothing kernel.  In
Section~3 we use these results to suggest the form that the initial
conditions should take, and show some of the problems that can arise
if these are not chosen carefully.  In Section~4 we describe the
major weakness of \SPH\ --- its inability to cope well with boundaries.
In Section~5 we suggest a list of diagnostic tests that help with the rapid
identification of errors, and to conclude we provide a summary of
basic prerequisites, initial conditions, tests and checks which, if
followed, should help to avoid, or at least trap, some of the problems
that we have encountered while using \SPH.

\section{Numerical Aspects of \SPH}

Before discussing the application of \SPH\ to physical systems we first
summarise some of the numerical implications of using a fully
Lagrangian particle code.

The essentials of the \SPH\ method can be
expressed in just two ideas:
\begin{enumerate}
\item The properties of the fluid at any point are estimated by taking
      a weighted average of those properties over the surrounding volume.
\item The continuum is approximated by a finite number of particles
      which are free to move under their mutual interaction and the
      action of any external body forces.
\end{enumerate}
Expressed mathematically, the first point is written
\begin{equation}
                \tilde{q}\left( {\bf r} \right)
    = \int_{0}^{\tau} q \left( {\bf s} \right) W \left( x \right) d^{3}\!s
\end{equation}
with
\begin{equation}
                x = \left| {\bf s} - {\bf r} \right| .
\end{equation}
Here $q$ is some property of the fluid, with the tilde denoting an
estimated value.
$W\left( x \right)$ is the weighting, or smoothing, function,
and the integral is taken over a volume $\tau$ about ${\bf r}$.
The second point above implies that the integral in equation~(1) is to be
replaced by a summation over particles within the volume $\tau$
(which are labelled by the subscript $i$),
\begin{eqnarray}
                \tilde{q} \left( {\bf r} \right) & \simeq &
                \sum_{i=1}^{n} q \left( {\bf s}_{i} \right)
                W \left( x_{i} \right) \Delta V_{i}  \\
                \Delta V_{i} & = & \frac{m_{i}}{\rho_{i}} .
\end{eqnarray}
If the smoothing kernel is centred on a particle, then we have ${\bf
r} \equiv {\bf r}_{j}$.  For a random distribution of particles
equation~(3) is in fact a Monte Carlo estimate of the integral.  The
smoothing volume, $\tau$, is varied so that $n$ is kept approximately
constant and equal to some prescribed number, $N$, which is generally
between 30 and 70,
\begin{equation}
\tau = \frac{N}{\tilde{\rho} \left( {\bf r} \right) / m_{j}} ,
\end{equation}
and usually a smoothing length, $h$, is defined as equal to
half the radius of the smoothing volume (in three dimensions) or
smoothing area (in two dimensions), see equation~(14).
For a distribution of particles with a local average density $\rho_{0}$,
\begin{equation}
n = \tau \frac{\rho_{0}}{m_{i}} = N \frac{\rho_{0}}{\tilde{\rho}} .
\end{equation}
Hereafter, particles are assumed to have unit mass.

We see then that there are two distinct sources of error within an \SPH\
simulation.
Systematic errors arise because the averaging process implied by equation~(1)
inevitably smooths out spatial variation in the fluid so that in
general $\tilde{q} \left( {\bf r} \right) \neq q \left( {\bf r} \right)$.
Furthermore $\tilde{q}$ will be a biased estimate if the smoothing
function, $W$, is incorrectly normalised.
Secondly, summing over a finite number of particles leads to statistical
error, dependent upon both the distribution of the particles and
the form of the smoothing function.
Before discussing these errors in turn, it is first necessary to
describe the behaviour of \SPH\ particles.

\subsection{Particle Distribution}

The particles of an \SPH\ simulation do not sit randomly, but are
distributed by dynamical forces in a manner analogous to the behaviour
of molecules in a real gas.  A system of particles initially
positioned at random will, over a few time steps, tend to adopt a more
uniform distribution as the repulsive force between neighbours acts to
inhibit close encounters and maximise the average spacing.  This
adaptive nature of \SPH\ is one of its strengths, but there are
significant consequences that arise from the properties of this \SPH\
gas.  The particle distribution is of particular importance since the
fluid is not sampled at random, but rather the smoothing kernel is
always centred on a particle.  This reinforces any effects that are
due to a correlation in the relative positions of the particles.

The \SPH\ gas can adopt one of three distinct states: chaotic,
thermalised (or relaxed), and crystalline.

\subsubsection{Chaotic Gas}

By chaotic we mean a system with particles positioned at random and
completely independent of each other.  Thus the probability density
function, $p$, (where $dn = \rho_{0} p \, d \tau$) is
\begin{equation}
 p \left( {\bf r},x \right) = \mbox{const} .
\end{equation}
This distribution function is shown in \fig\ref{chaotic}.  This is
the situation envisaged in early discussions of \SPH\
(e.g.\ Lucy
[1977], Gingold \& Monaghan [1977], Hernquist \& Katz [1989]).
The chaotic gas however is not in dynamic equilibrium and will evolve in
time, i.e.\ become thermalised.

\begin{figure}
 \centering
 \caption{Probability density function for a chaotic gas}
\label{chaotic}
\end{figure}

\subsubsection{Thermalised Gas}

By a thermalised, or relaxed, \SPH\ gas we mean a system of particles
that is in dynamic equilibrium.  In this state the velocity
distribution of the particles is Maxwellian, as would be expected for
a real gas.  The probability density function $p \left( {\bf r}_{j},x \right)$
centred on a particle, $j$ (which is not itself included in the calculation
of $p$) is constant at large separations
(typically one smoothing length or more), but is zero near the
particle and rises to a peak slightly in excess of the average
before dropping to the constant level. This distribution function is shown
in \fig\ref{thermal}.
Note that this effect is due to a correlation between particles;
for a randomly selected point the constant distribution~(7) is
recovered.  In all that follows (and in \SPH\ generally) we
shall assume the smoothing kernel to be centred on a particle.

\begin{figure}
 \centering
 \caption{Probability density function for a thermalised gas}
\label{thermal}
\end{figure}

\subsubsection{Crystalline Structure}

If the kinetic energy of the gas is reduced slowly to zero, a
crystal-like structure will appear as the repulsive nature of the
inter-particle force encourages particles to adopt maximum
separations.  Theoretically $p \left( {\bf r}_{j},x \right) $ will
reduce to a series of delta functions corresponding to either 8 or 16
lattice points intersecting the volume element, but in practice flaws,
dislocations and slight displacements from an exact grid mean that $p
\left( x \right)$ shows two or three sharply defined peaks before
again tending to a constant value at larger separations, (see
\fig\ref{crystal}).  In the presence of a potential field, particles
will tend to form lines or layers along equipotentials.

\begin{figure}
 \centering
 \caption{Probability density function for a crystalline gas}
\label{crystal}
\end{figure}

In this state numerical effects dominate as the movement of particles
is severly restricted, although the statistical errors will be at a
minimum.  This will be true whenever the average kinetic energy
approaches zero.

\subsection{Error Analysis}

The systematic errors associated with \SPH\ calculations can be put into
three broad classes: errors dependent upon the distribution of the
particles, errors in estimated quantities due to the smoothing kernel
always being centred on a particle, and errors arising from the
presence of a density gradient.  All the systematic errors are
dependent upon the shape of the smoothing kernel.

We stated above that equation~(3) is a Monte Carlo estimate of the
integral given in equation~(1), but this is strictly true only for a
chaotic distribution of particles when $p \left( {\bf r}_{j},x \right)$
is constant.  In general equation~(3) is actually an estimate of the
integral
\begin{equation}
\theta = n \left( {\bf r}_{j} \right) \int_{0}^{\tau}
   \frac{ q \left( {\bf s} \right) }{\rho \left( {\bf s} \right) }
      W \left( x \right) p \left( {\bf r}_{j},x \right) d^{3}\!s ,
\end{equation}
where $n$ is the number of neighbouring particles.  As $W$ is
normalised such that $\int_{0}^{\tau} W d^{3}\!s = 1$, we can define a
bias parameter
\begin{equation}
\beta =
   \int_{0}^{\tau} \left( W-1 \right) p\left( {\bf r}_{j},x \right) d^{3}\!s .
\end{equation}
$\beta$ is zero for a chaotic distribution, but becomes negative for a
relaxed \SPH\ gas and centrally peaked kernel when $p$ tends to zero for
small $x$.

There is a further, and more important, consideration.  The smoothing
kernel is always centred on a particle, which means that as well as
the $n$ particles that give the Monte Carlo summation there is always
an additional contribution from the central particle itself.  This
then gives an expected value of $\tilde{q}$ (for a uniform gas with $q
\left( {\bf r} \right) = q_{0}$ and $\rho = \rho_{0}$ throughout) of
\begin{equation}
\varepsilon \left( \tilde{q} \right)
   = \frac{q_{0}}{N} \left\{ W \left( 0 \right) \tau
     + n \left( \beta + 1 \right) \right\} .
\end{equation}
Clearly, for $\beta=0$, a more centrally peaked smoothing kernel gives
a larger expectation.  For the most commonly used Beta-spline kernel
(which is highly peaked about $x=0$
, see equation~[14]) and $N=32$, this implies an over-estimate of about
18\% and 33\% for a chaotic (un-biased) \SPH\ gas in two and three
dimensions respectively.  On subsequent steps, if the particles
were to remain fixed, the error worsens as $q_{0}$ is updated to $\tilde{q}$.
{}From equation~(10), the equilibrium expectation of density with $\beta = 0$
is approximately 22\% and 50\% greater in two and three dimensions than
the true value.
In practice, however, the particles will begin to redistribute
themselves before this equilibrium is reached, creating a negative
bias and reducing the overestimate.
Note that \SPH\ is an iterative process with two relevent timescales.
Particles move toward dynamic equilibrium and there is the numeric
equilibrium of equation~(10).

The statistical error associated with a Monte Carlo estimate
of a function can be determined from the variance of that function
(see, for example, Hammersley \& Hanscomb 1964).
We have
\begin{equation}
\tilde{q} = \varepsilon \left( q \right) \pm \frac{\sigma}{\sqrt{n}}
\end{equation}
where
\begin{equation}
\sigma^{2}
 = \int_{0}^{\tau} \left( W - \theta \right)^{2} p\left( x \right) d^{3}\!s .
\end{equation}
Equation~(12) only holds when the particles are independent of each other.
For a relaxed \SPH\ gas there is a strong negative correlation in the
particle positions which significantly reduces the statistical error.
A chaotic distribution of particles has a statistical error of about
20\%, but this drops rapidly to less than 10\% as the \SPH\ gas becomes
thermalised. It is therefore important that the particle distribution
stays thermalised throughout the simulation.

The distribution function $p \left( {\bf r},x \right)$ is a function not
only of the dynamic state of the particles, but also of the non-random
density gradient and any non-linear variation of the large scale
density will be incorrectly estimated by equation~(3).  $\tilde{q} \left(
 {\bf r} \right)$ will be over-estimated whenever $\rho'' \left( {\bf r}
\right)$ is positive and under-estimated for $\rho'' \left( {\bf r}
\right)$ negative (any symmetrical smoothing of an odd function gives
a value correct at the centre of the kernel).  If the density over an
extended region can be approximated by
$\rho = \rho_{0} \left( 1 + bx + cx^{2} \right)$
then for the most commonly used Beta-spline kernel,
\begin{equation}
  \tilde{\rho} = \rho_{0} \left( 1 + \alpha . c h^{2} \right)
\end{equation}
where $\rho_{0}$ is the correct value, h is the smoothing length,
and $\alpha = 31/98$, 31/105 in two and three dimensions respectively.
The size of this error is again dependent upon the shape of the
smoothing kernel, but now a more centrally peaked kernel gives a smaller
error.  The dependence on the smoothing length simply shows that
more particles are needed to follow a more rapidly changing
density gradient, and equation~(13) can be used to check that
the resolution within a simulation is sufficient to follow
the change in density.

An error in the density calculation clearly introduces further errors
into the inter-particle force, but in addition to this there are two
other factors that reduce the accuracy of the force calculation.
Firstly, as pointed out by Hernquist \& Katz (1989) and others, two
particles with different densities will each have a different
smoothing length.  This may lead to a non-symmetrical application of
the force between them -- the particle with a lower density and larger
smoothing length will see a higher density neighbour near the edge of
its kernel, but then will not be seen by that neighbour in turn.  As
this only affects neighbours at the edge of the kernel, the errors
will be small but they can become important in regions with a large
density gradient.  For this reason a symmetric force law should
be used Monaghan (1992).

The second error is a little harder to quantify.  It is assumed that
the \SPH\ particles mark the centre of mass of an element of gas.
This may not always be the case, particularly in regions of steep
density gradients.  Although the gas attached to an \SPH\ particle is
never referred to explicitly, the local state of the fluid, as
determined by the \SPH\ simulation, must be represented in the gas
around each particle.  In particular a density gradient in the global
fluid, as described by the particle distribution, must be assumed to
exist also across the local or `virtual' fluid associated with a
particle.  This means that as the density gradient around a particle
changes, either because the particle moves from one region to another
or because the local gradient is changing with time, the centre of
mass of this fluid element will move and will no longer coincide with
the particle.  It follows that the force between fluid elements, which
is applied between particles, is now being applied at the wrong
position.  Note that, since the fluid around a particle is never
described explicitly, this error can only appear when an individual
particle experiences a change in the density gradient. This is a
subtle error that we have found difficult to quantify. It leads to no
obvious problems for the models we have applied \SPH\ to but is an
inherent feature of the \SPH\ formalism.

It is also important to ensure that the simulation uses enough
particles to model all the structure that makes up the physical system.
If the code has a maximum smoothing length, then the number of
neighbours can fall far below $N$ in low density regions.
In the case of a convective stellar envelope, the envelope must be at
least several smoothing lengths deep to have any hope of following
vertical transport processes.  As shown above, more particles are
needed to resolve a large density contrast.
But perhaps the most important consideration is the inclusion of
3-dimensional shocks. Thomas \& Couchman (1992)  have shown that at
least 1000 particles are needed to follow a thermalised shock region
adequately in 3-dimensions, which implies, for instance, that to
follow the supersonic collision of two gas clouds will require tens of
thousands of particles in each cloud.

\subsection{The Smoothing Kernel}

The expected value of $\tilde{q}$ and the numerical error are both
sensitive functions of the smoothing function, $W$.  A more centrally
peaked function gives a larger bias for a chaotic particle
distribution, but a smaller error in regions with high density
gradient.  A kernel falling to zero at the centre would give the
correct estimate in these situations, but a centrally condensed kernel
has other advantages.  In particular, weighting the central particle heavily
in effect smooths in {\em time} as large fluctuations from one time step
to the next are prevented.

The most commonly used smoothing function is the Beta-spline kernel,
due to Monaghan \& Lattanzio (1985)
\begin{equation}
W \left( {\em r},h \right) = W_{0}  \left\{
 \begin{array}{ll}
                        4 - 6x^{2} + 3x^{3}  &   0 \leq x < 1 \\
                        \left( 2 - x \right)^{3} & 1 \leq x < 2 \\
                        0                        & x \geq 2
 \end{array} \right.
\end{equation}
where $x = r/h$ and $W_{0} = 5/14\pi h^{2} = 10/7 \tau$ in two
dimensions, and $W_{0} = 1/4 \pi h^{3} = 8/3 \tau$ in three
dimensions.  This is very centrally condensed and falls smoothly to
zero at the outer edge.  Extensive tests by several authors, (Gingold
\& Monaghan 1983, Monaghan \& Lattanzio 1985, Pearce 1992),
have shown the $\beta$-spline kernel to be the superior to other
possible choices.  In particular, any kernel without compact support
or that is not smooth (e.g.\ top-hat, triangular gaussian cut-off)
should all be avoided as they introduce severe numerical effects
(ringing, density overestimates and interpenetration of particles) and
are unable to model shocks accurately.

$W'({\em r},h)$ is needed for the force calculation, but $W'$ can
be found from a different kernel to that used for the smoothing
with no loss of consistency.
The inter-particle force given by equation~(15) falls to zero at
$x=0$ which under some circumstances can encourage particles
to form clumps, giving large numerical errors.  To inhibit the
formation of clumps of particles, and to force particles to adopt
a more uniform spacing, which reduces the statistical error,
Thomas \& Couchman (1992) use a second kernel for their force law
which gives a constant inter-particle force at small separations,
\begin{equation}
W \left( {\em r},h \right) = W_{0}
  \left\{ \begin{array}{ll}
                     \frac{44}{9} - 4x    &   0 \leq x < \frac{2}{3} \\
                     4 - 6x^{2} + 3x^{3}  &   \frac{2}{3} \leq x < 1 \\
                     \left( 2 - x \right)^{3} & 1 \leq x < 2 \\
                     0                       & x \geq 2
  \end{array} \right. .
\end{equation}
Using this kernel for the force law, a chaotic \SPH\ gas relaxes
about 10\% faster.
\section{Initial Conditions}

For any integration scheme used to follow the evolution of a physical
system in time it is clearly important that the boundary conditions at
$t=0$ are well defined.  Only then can one know precisely which
problem is being solved and in what way the final results
are a function of the initial conditions.

For a time dependent SPH simulation, this means that the initial
conditions must be relaxed.  It is not sufficient to lay particles
down at random to create the desired density profile and then supply
the relevant temperatures and velocities, for in this case there will
be jumps in density, entropy, kinetic energy and the total energy of
the system (from the release of binding energy).
The jumps in smoothed quantities are caused both by changes in the
particle distribution as the system moves toward dynamic equilibrium
(thereby changing the distribution function, and hence the bias), and
also by changes in the SPH estimate independent of the distribution
(equation [10]).

\begin{figure}
 \centering
 \caption{Evolution of the mean density in a wrapped 2D box}
\label{densevol}
\end{figure}

Even with boundary conditions held constant, it can take many ($\simeq
50$) time steps before the particles reach dynamic equilibrium.  This
is illustrated in \fig\ref{densevol} where
8000 particles with a uniform temperature were placed at random in
a fully wrapped 2-D box.  For the first two time steps, before the
particles have time to move, the initial average density is close to 1.18
times the true value and rises to 1.25, as predicted by equation~(10).
The statistical error is in excess of 25\% (lines of $1 \sigma$ are
plotted).
Once the particles start to move, the SPH gas begins to relax and the
bias quickly becomes negative.  This tends to compensate for the
particle centred kernel, and the average density then approaches (but
does not quite reach) the true value, settling about 1\% too high.
By the 50th time step, the particles have reached dynamic equilibrium
and the oscillations (due to sound waves crossing the box) have nearly
vanished.  The statistical error is now less than 10\% and the
particle distribution is as shown in \fig\ref{thermal}.  The qualitative
behaviour is identical in three dimensions, but the particles relax
more quickly ($\simeq 30$ time steps).

Ideally, the particles for an SPH simulation should be allowed to
relax with the boundary conditions held constant before the simulation
proper is begun.  Note that there is a steady decrease in the average
kinetic energy due to the artificial viscosity, which has a damping
effect.  Care must be taken to ensure that the particles do not become
over-damped, when the lattice effects described in Section~2.1.3 will
begin to dominate their motions.

An alternative approach to relaxing particles in situ is to relax the
particles within a regular cube and then transform them into the
required geometry.  However, compression in one direction only does
not work as neither dynamic or numerical equilibrium is maintained.
Reasonable results for a spherical system can be obtained by ordering
the particles in distance from a fixed point and then transforming
this distance to obtain the desired density profile. This preserves
the spatial correlations of the relaxed box but still leads to some
residual spurious density enhancement.

Transforming particles from a uniform grid with no random
placement or relaxation into a non-cartesian gravitational potential
gives the worst possible starting conditions, introducing many
numerical artifacts.  The strange features seen in a plot of returned
SPH density against radius for one of these transformations is shown
in \fig\ref{strange}. The artificial density enhancements are now
correlated in space, producing hot and cold regions a few steps after
the start.

\begin{figure}
 \centering
 \caption{Strange density artifacts obtained from grid like initial conditions}
\label{strange}
\end{figure}

\section{Edge Effects}

\SPH\ copes badly with any edge or boundary within a simulation, whether
the boundary is `natural' (e.g.\ the edge of a gas cloud where the
density falls to zero) or artificial (e.g.\ the inner high density
edge of a stellar envelope when only a fraction of a star is to be
modelled).

A sharp edge in the initial conditions (e.g.\ if the density profile
of a gas cloud is cut off at some finite radius) will expand from the
first time step, and also send a sound wave inwards.  This can be
important if the inward moving sound wave is focussed and amplified.
It is best to keep the majority of the particles unaware that an outer
edge exists by placing the low density cut-off at a radius far beyond
the regions of interest and keeping track of how far the edge effects
propagate inwards during the simulation.

Artificial boundaries come with their own set of problems.  As an
illustration, take gas in a box which initially is completely wrapped
so that the gas is unbounded.  An ideal artificial boundary could be
inserted into the box, or used to replace one of the wrapping walls,
without changing the distribution or average quantities of the \SPH\
particles.  A brick wall, or mirror, is not good enough.  Particles
approaching the wall will not see any neighbours on the other side and
will be rapidly accelerated towards the wall (remembering that $\nabla
P / P$ within the gas can be very small) leading to a large density
excess, heating, and slowing the code down as high velocities and
accelerations need small time steps.

If the state of the gas is known at the boundary (i.e.\ the properties
of the gas are given there as a boundary condition) then one of three
methods could be considered.  i)~Particles could be placed on the far
side of the boundary to a depth of two smoothing lengths and held
fixed to maintain the desired density and temperature profile.  There
are several difficulties with this approach.  There is nothing to
prevent a particle from penetrating the boundary, and once through a
particle will again have insufficient neighbours.  Various numerical
catches can be thought of to overcome this, but none of them is
entirely satisfactory.  It is also very difficult to create the
correct distributions in the particle properties, in other words it is
difficult to set up the boundary particles to look relaxed to
neighbours on the other side of the wall.  Because the boundary
particles are fixed, they cannot react to any change in the simulation
particles near them, removing one of the main strengths of
\SPH\ --- its ability to adapt.  This method can also be
inefficient with computing time if the boundary lies in a high density
region, as many particles will have to be used which are not
contributing anything to the simulation.

ii) An alternative approach is to use an analytical evaluation of
smoothed quantities, and is perhaps the simplest method to adopt if
conditions allow.  The smoothing kernel of a particle near the
boundary will overlap the boundary.  If the properties of the gas are
given at the boundary, then the contribution from this overlap region
can be calculated analytically (or numerically and tabulated) and
added to the numerical sum over neighbours.  Because this is really
\SPH\ in the limit of an infinite number of particles, it overcomes
the difficulty of finding relaxed conditions, and is computationally
efficient as no extra particles are needed.  But again, this gives a
fixed boundary that cannot adapt to changing conditions on the
particle side, and there is nothing to stop penetration by
particles.  Furthermore, in non-symmetric geometries where the
boundary is neither planar or spherical, or does not lie along a
surface with constant gas properties, it may be difficult to find the
analytical function for a general point near the boundary.

iii) For simulations involving a gravitational potential which is
independent of the particles, the problems covered above can be
overcome by removing the high density edge altogether.  Below the
region of interest the physical potential, $\Phi \left( {\bf r}
\right)$, is replaced by an artificial potential, $\Psi \left( {\bf r}
\right)$, which is set to rise to the surface value within
some short distance ($\approx$ one smoothing length) of the boundary.
Now the only edges seen by the particles are zero density ones, where
the small number of particles have limited impact upon the majority
which move freely within the potential well that has been created.

This method has been used successfully to provide a lower boundary
when modelling the common envelope enclosing a contact binary system
(Martin 1993).  Below the lower boundary, the effective
gravity is calculated from an artificial potential that is taken to be
a function of the real potential, $\Psi = \Psi \left( \Phi \left( {\bf
r} \right) \right)$, and matched onto $\Phi$ at the boundary so that
the first three derivatives are continuous.  In this way the effective
gravity remains parallel to the real value while smoothly changing
direction.  This approach has been found to give acceptable results in
spite of the rapid change in the effective gravity and the enormous
density gradients within the artificial potential region.

In general, edge-effects can lead to serious problems if they are not
closely monitored, but if the edge is placed suitably far beyond the
region of interest then the inaccuracies introduced can be
limited, although this will often be at a high computing cost as many
particles not directly relevant to the simulation must be used.
\section{Diagnostics}

\SPH\ codes, like any large computer program, are susceptible to the
FURLRATS condition (`fiddle until results look right and then stop').
To help overcome this, we suggest that the following short list of
basic tests and checks be applied to every \SPH\ simulation.
Experience has shown that not only do almost all of the errors (that
we know about) show up in one or more of these checks, but also by
having all this information available for every run it is often
possible to narrow down the probable source of the error and save many
hours of debugging time.  We divide these tests into three broad
categories.

\subsection{Global Checks}

Most importantly, conserved quantities (mass, total energy, linear and
angular momentum) should be examined at every time step.  Momentum
will be conserved exactly, as will mass unless the density is
calculated from the velocity divergence (see Monaghan 1992,
\S 3.2).  The total energy should be conserved to whatever accuracy is
required.  Significant jumps in any of these quantities implies
that something is wrong with the code. The most likely source of
error is attempting to use too long a time step.

Follow the average density, total entropy and total kinetic energy of
the particles.  A sudden change in any of these quantities in the
first few time steps shows that the initial conditions were not
relaxed.  This is a fatal flaw since any conclusions based on the
evolution of initial conditions that were not relaxed will be
unreliable.  As it is almost impossible to lay particles down in a
perfectly relaxed state, we suggest that the initial condition should
be evolved for a few time steps and then this distribution used for
all later comparisons.

\subsection{Particle Checks}

The distribution of particle properties should be monitored closely;
if a particle has extreme values in any variable when compared to
the majority, then something is wrong!
Errors in the code that cause a Single Particle Discrepancy are generally
the most difficult to locate, but often have the most far reaching
consequences.  Ignore that spurious particle at your peril.

Entropy is a particularly useful quantity to follow as it shows up
changes in density and temperature simultaneously.  A large spread in
entropy can indicate a problem.  In a gas sphere low entropy gas will
sink whilst high entropy gas rises and any scatter in entropy leads to a
gradual segregation and a systematic drift in the sphere's properties.
The entropy of each particle is conserved for an ideal gas in the
absence of shocks, and the change in entropy of individual particles
should also be examined.  Spurious jumps in a particle's entropy
outside of shocks is frequently the result of too large a time step.
If particles move too far on a single time step then close neighbours
are created. This moves the particle distribution away from the
thermalised condition to the chaotic one and, as we have shown, \SPH\
does not work very well in this regime.

\subsection{Numerical Checks}

We also suggest following the average number of neighbours.  This
gives an indication of the difference between the real particle
density and the \SPH\ density, (equation~[6]), as well as
confirming that sufficient neighbours are being found.  If the average
number of neighbours is changing rapidly then the code is probably far
from convergence.  Check also that the \SPH\ gas is not becoming
crystalline.A useful diagnostic of the code efficiency can be obtained by
monitoring the number of particles tested for being a neighbour
compared to the number actually found. This ratio should be small in
an efficient scheme. We think a value of 150 is reasonable.

Using \SPH\ with a variable smoothing length would seem to allow small
numbers of particles to detach from the system as a whole.  Consider
100 particles in a very dense, cold clump. They see each other but
have very little knowledge of the diffuse gas surrounding them.  If
cooling is implemented then it would appear to be possible for this
clump to evolve entirely independently. Moral: don't trust small
groups of isolated particles whose properties are far from those of
the surrounding gas.  The only way to handle groups of isolated
groups properly is to use more particles.

Finally, the code should always be run for several test problems with
known solutions (e.g.\ shock tubes, static gas cloud configurations),
and if possible results confirmed using first a different number of
particles and then a shorter time step.  To get reliable results you
must ensure that the \SPH\ gas remains in the thermalised regime
throughout the run, especially during the first few steps.
\section{A try-before-you-fly list for first-time \SPH\ programers}

\begin{itemize}
\item Refer to Monaghan (1992) for \SPH\ equations.
\item Use the $\beta$-spline kernel (Monaghan \& Lattanzio 1985), equation~15,
      anything else just isn't good enough.
\item Use a variable smoothing length.
\item Use a symmetrical force law (gives equal and opposite force
      directly so conserves momentum and also overcomes systematic
      error due to variable smoothing length).
\item Ensure particles are relaxed at $t=0$; never place particles
      on a grid.
\item In 3-dimensions the total number of particles should be in
      excess of 10,000 when shocks are present. If only global
      properties are being examined then fewer particles are required.
      Try repeating your simulations with fewer particles to see
      if the results are altered.
\item Repeat all or part of the simulation with a different time step.
      A shorter time step should not change the results.
\item \SPH\ works best in the thermalised regime. Check that the gas
      is not spending time in either the crystalline or chaotic states.
\item Monitor the scatter in entropy induced by your code. Too large
      a spread will introduce temperature gradients into your results.
\item Check the code on problems with known solutions.
\end{itemize}

\section{Conclusions}

\SPH\ has many advantages over grid based codes, in particular the fully
Lagrangian treatment makes working with non-symmetric geometries
straight forward, and the absence of a pre-determined grid allows
evolution from the initial conditions as well as overcoming all the
numerical problems that can occur with a grid structure.  \SPH\ is
also remarkably robust, producing accurate results over a huge range
of conditions with parameters spanning many orders of magnitude.

As we have shown, \SPH\ comes complete with its own set of
difficulties, and thought must be put into choosing initial conditions
and following the simulation to ensure that the results are accurate
and relevant.  \SPH\ behaves is an iterative process, and large
numerical errors can arise whenever the \SPH\ gas is far from
equilibrium.  Numerical errors will also be significant whenever there
are insufficient particles to model the physical structure, (\eg
shocks or regions of sharply changing density gradients), to a high
enough resolution.  \SPH\ copes badly with any boundary, requiring
extra care whenever one is present.

However, if sufficient care is taken and the suggestions given here
followed, then we can guarantee you hours of fun and enjoyment with
your code.

\section*{Acknowledgements}

TJM was supported by the DSS.
We acknowledge the facilities of the STARLINK minor node at Sussex.

\section*{References}
\parindent=0pt\parskip=4pt\hangindent=3pc\hangafter=1

Benz, W., Cameron, A. G. W., Melosh, H. J., 1989 Icarus, 81, 113

Binney, J., Tremaine, S., 1987, Galactic Dynamics.
    Princeton University Press, Princeton

Evrard, A. E., 1990, \ApJ , 363, 349

Gingold, R. A., Monaghan, J. J., 1977, \MN , 181, 375

Gingold, R. A., Monaghan, J. J., 1983, \MN , 204, 715

Hammersley, J. M., Handscomb, D. C., 1964, Monte Carlo
Methods. Methuen

Hernquist, L., Katz, N., 1989, \ApJS , 70, 419

Lanzafame, G., Belvedere, G., Molteni, D., 1993, \MN , 263, 839

Lucy, L. B., 1977, \AJ , 82, 1013

Martin, T. J., 1993 \PhD , University of Sussex

Meglicki, Z., Wickramasinghe, D., Bicknell, G. V., 1993, \MN , 264, 691

Monaghan, J. J., 1992, \ARAA , 30, 543

Monaghan, J. J., Lattanzio, J. C., 1985, \AaA , 149, 135

Navarro, J., S. D. M. White, 1993, \Pre , University of Durham

Pearce, F. R., 1992, \PhD , University of Sussex

Pearce, F. R., Thomas, P. A., Couchman, H. M. P., 1993, \MN , \Pre ,
University of Sussex

Pongracic, H., Chapman, S. J., Davies, J. R., Disney, M. J.,
Nelson, A. H., Whitworth, A. P., 1992, \MN , 256, 291

Steinmetz, M., M\"{u}ller, E., 1993, \AaA, 268, 391

Thomas, P. A., Couchman, H. M. P., 1992, \MN , 257, 11

\end{document}